\documentclass[twoside,11pt]{article}

\usepackage[final]{melba}

\usepackage[utf8]{inputenc} 
\usepackage[T1]{fontenc}    
\usepackage{hyperref}       
\usepackage{url}            
\usepackage{booktabs}       
\usepackage{amsfonts}       
\usepackage{nicefrac}       
\usepackage{microtype}      
\usepackage{natbib}
\setcitestyle{square}
\usepackage{graphicx}
\usepackage{subfigure}
\usepackage[table]{xcolor}
\usepackage{float}
\usepackage{multirow}

\usepackage{caption}
\usepackage{enumitem}
\usepackage{listings}
\usepackage{xcolor}

\definecolor{codegreen}{rgb}{0,0.6,0}
\definecolor{codegray}{rgb}{0.5,0.5,0.5}
\definecolor{codepurple}{rgb}{0.58,0,0.82}
\definecolor{backcolour}{rgb}{0.95,0.95,0.92}

\lstdefinestyle{mystyle}{
    backgroundcolor=\color{backcolour},   
    commentstyle=\color{codegreen},
    keywordstyle=\color{magenta},
    numberstyle=\tiny\color{codegray},
    stringstyle=\color{codepurple},
    basicstyle=\ttfamily\footnotesize,
    breakatwhitespace=false,         
    breaklines=true,                 
    captionpos=b,                    
    keepspaces=true,                 
    numbers=left,                    
    numbersep=5pt,                  
    showspaces=false,                
    showstringspaces=false,
    showtabs=false,                  
    tabsize=2
}

\hypersetup{
    colorlinks,
    citecolor=gray,
    linkcolor=black,
    urlcolor=black
}

\melbaheading{2}{https://www.melba-journal.org/article/18272-covid-19-image-data-collection-prospective-predictions-are-the-future}{2020}{1-38}{07/2020}{12/2020}{Cohen, Morrison, Dao, Roth, Duong, Ghassemi}{}{}

\ShortHeadings{COVID-19 Image Data Collection: Prospective Predictions are the Future}{Cohen, Morrison, Dao, Roth, Duong, Ghassemi}
\firstpageno{1}

\title{COVID-19 Image Data Collection: \\Prospective Predictions are the Future}

\author{%
  Joseph Paul Cohen \email joseph@josephpcohen.com \\
  \addr Mila, University of Montreal
  \AND
  Paul Morrison \email paul.j.f.morrison@gmail.com\\
  \addr Mila, Fontbonne University
  \AND
  Lan Dao \email phuong.lan.dao@umontreal.ca\\
  \addr Department of Medicine \\
  Mila, University of Montreal
  \AND
  Karsten Roth \email karsten.rh1@gmail.com\\
  \addr Vector, Mila, Heidelberg University
  \AND
  Tim Q Duong \email tim.duong@stonybrookmedicine.edu \\
  \addr Stony Brook Medicine
  \AND
  Marzyeh Ghassemi \email marzyeh@cs.toronto.edu\\
  \addr Vector, University of Toronto
}

\def\blind#1{#1}

\begin{document}

\maketitle

\begin{abstract}
Across the world's coronavirus disease 2019 (COVID-19) hot spots, the need to streamline patient diagnosis and management has become more pressing than ever. As one of the main imaging tools, chest X-rays (CXRs) are common, fast, non-invasive, relatively cheap, and potentially bedside to monitor the progression of the disease. This paper describes the first public COVID-19 image data collection as well as a preliminary exploration of possible use cases for the data. This dataset currently contains hundreds of frontal view X-rays and is the largest public resource for COVID-19 image and prognostic data, making it a necessary resource to develop and evaluate tools to aid in the treatment of COVID-19. It was manually aggregated from publication figures as well as various web based repositories into a machine learning (ML) friendly format with accompanying dataloader code. We collected frontal and lateral view imagery and metadata such as the time since first symptoms, intensive care unit (ICU) status, survival status, intubation status, or hospital location. We present multiple possible use cases for the data such as predicting the need for the ICU, predicting patient survival, and understanding a patient's trajectory during treatment. 
Data can be accessed here: \blind{\url{https://github.com/ieee8023/covid-chestxray-dataset}}

\end{abstract}

\section{Introduction}

In the times of the rapidly emerging coronavirus disease 2019 (COVID-19), hot spots and growing concerns about a second wave are making it crucial to streamline patient diagnosis and management. Many experts in the medical community believe that artificial intelligence (AI) systems could lessen the burden on hospitals dealing with outbreaks by processing imaging data \citep{10.1007/s00330-020-06748-2,10.1148/radiol.2020201365}. Hospitals have deployed AI-driven computed tomography (CT) scan interpreters in China \citep{Simonite2020} and Italy \citep{Lyman2020}, as well as AI initiatives to improve triaging of COVID-19 patients (i.e., discharge, general admission, or ICU care) and allocation of hospital resources \citep{Strickland2020,Hao2020MITReview}. 

Carefully curated and annotated data is the first step to developing any diagnostic or management tool. While there exist large public datasets of more typical chest X-rays (CXR) \citep{WangNIH2017,Bustos2019PadChest,Irvin2019CheXpert,Johnson2019mimic-cxr, Demner-Fushman2016}, there was no public collection of COVID-19 CXR or CT scans designed to be used for computational analysis. We first made data public in mid February 2020 and the dataset has rapidly grown  \citep{Cohen2020coviddataset}. More recently, in June, the BIMCV COVID-19+ dataset \citep{DeLaIglesiaVaya2020BIMCVCOVID-19} was released. While it has more samples than the dataset we present, we complement their work with a focus on prospective metadata from multiple medical centers and countries.

Many physicians remain reluctant to share their patients' anonymized imaging data in open datasets, even after obtaining consent, due to ethical concerns over privacy and confidentiality and a hospital culture that, in our experience, does not reward sharing \citep{Keen2013, Lee2017medicaldata, Kostkova2016, Floca2014}.
In order to access data in one hospital, researchers must submit a protocol to the hospital's institutional review board (IRB) for approval and build their own data management system. While this is important for patient safety, such routines must be repeated for every hospital, resulting in a lengthy bureaucratic process that hurts reproducibility and external validation. 

The ultimate goal of this project is to aggregate all publicly available radiographs, including papers and other datasets that can be combined. For research articles, images are extracted by hand, while for websites, such as Radiopaedia and Eurorad, data collection is partially automated using scrapers that extract a subset of the metadata while we hand review case notes to determine clinical events. 
This work provides three primary contributions: 
\begin{itemize}
    \item We create the first public COVID-19 CXR image data collection, currently totalling 679 frontal chest X-ray images from 412 people from 26 countries and growing. The dataset contains clinical attributes about survival, ICU stay, intubation events, blood tests, location, and freeform clinical notes for each image/case. In contrast to other works, we focus on prospective metadata for the development of prognostic and management tools from CXR. Images collected have already been made public and are presented in an ML-ready dataset with toolchains that are easily used in many testable settings.
    \item We also present and discuss clinical use-cases and propose ML tasks that may address these clinical use cases such as predicting pneumonia severity, survival outcome, and need for the intensive care unit (ICU). Benchmark results using transfer learning with neural networks are included. 
    \item We also discuss how to use the location information in this dataset for a  Leave-One-Country/Continent-Out (LOCO) evaluation to simulate domain shift and provide a more robust evaluation.
\end{itemize} 
    
Currently, all images and data are released under the following GitHub URL\footnote{\blind{\url{https://github.com/ieee8023/covid-chestxray-dataset}}}. We hope that this dataset and tasks will allow for quick uptake of COVID-related prediction problems in the machine learning community. 

\section{Background and Related Work}

\subsection{Learning in Medical Imaging}

In recent years, ML applications on CXR data have seen rising interest, such as lung segmentation \citep{Gordienko2018LungSegmentation,islam2018robust}, tuberculosis and cancer analysis \citep{Gordienko2018LungSegmentation,stirenko_tuberculosis,Lakhani2017Tuberculosis},  abnormality detection \citep{islam2017abnormality}, explanation \citep{Singla2020Exaggeration}, and multi-modality predictions \cite{Rubin2018LateralMultiView,Hashir2020Lateral}. With the availability of large-scale public CXR datasets created with ML in mind (e.g. CheXpert \citep{Irvin2019CheXpert}, Chest-xray8 \citep{WangNIH2017}, PadChest \citep{Bustos2019PadChest} or MIMIC-CXR \citep{Johnson2019mimic-cxr}), neural networks have even reported being able to achieve performance near radiologist levels \citep{Rajpurkar2018CheXNeXt,rajpurkar_chexnet:_2017,Irvin2019CheXpert,putha2018artificial,Majkowska2019,Putha2018CQ100k}. However, the adoption has its own challenges \citep{Couzin-Frankel2019medicalcontends}. 

\subsection{Use of Imaging for COVID-19}
In this section we will give a high level overview of the recommendations for the clinical use of imaging in COVID-19 patient management. It should be noted that this list is only meant to help guide algorithm development and is not medically comprehensive.

Ever since the dawn of the outbreak, imaging has stood out as a promising avenue of research and care \citep{10.1148/radiol.2020200490,10.1148/radiol.2020200847}. Particularly in the beginning of the outbreak, computed tomography (CT) scans captured the attention of both the medical \citep{10.1148/ryct.2020200034,10.1148/radiol.2020200527} and the ML \citep{10.1016/s2589-75002030054-6} communities.

From March to May 2020, the Fleischner Society \citep{10.1148/radiol.2020201365}, American College of Radiology (ACR) \citep{ACR2020COVIDImaging}, Canadian Association of Radiologists (CAR) \citep{CAR2020COVIDStatement},  Canadian Society of Thoracic Radiology (CSTR) \cite{Dennie2020-CSTR-CAR-Statement}, and British Society of Thoracic Imaging (BSTI) \citep{Nair2020bst} released the following recommendations:
\begin{enumerate}
\item Imaging tests (CXR and chest CT scans) should not be used alone to diagnose COVID-19 nor used systematically on all patients with suspected COVID-19;
\item Findings on CT scans and CXR are non-specific and these imaging techniques should not be used to inform decisions on whether to test a patient for COVID-19 (in other words, normal chest imaging results do not exclude the possibility of  COVID-19 infection and abnormal chest imaging findings are not specific for diagnosis); 
\item CXR and chest CT scans can be used for patients at risk of disease progression and with worsening respiratory status as well as in resource-constrained environments for triage of patients with suspected COVID-19.
\end{enumerate}

However, CXRs remain the first choice in terms of the initial imaging test when caring for patients with suspected COVID-19. CXRs are the preferred initial imaging modality when pneumonia is suspected \citep{ACR2018Appropriateness} and the radiation dose of CXR (0.02 mSv for a PA film) is lower than the radiation dose of chest CT scans (7 mSv), putting the patients less at risk of radiation-related diseases such as cancer \citep{FDA2017Radiation}.

In addition, CXR are cheaper than CT scans, making them more viable financially for healthcare systems and patients. \citep{10.4329/wjr.v7.i8.189, 10.21037/atm.2017.07.20}. 
Finally, portable CXR units can be wheeled into ICU as well as emergency rooms (ER) and are easily cleaned afterwards, reducing impact on patient flow and risks of infection \citep{CAR2020COVIDStatement, ACR2020COVIDImaging}.

A recent hypothesis that is increasingly confirmed by research suggesting that COVID-19 is not a pulmonary but rather a vascular disease \citep{varga2020endothelial} \citep{ackermann2020pulmonary}. This could explain the variety of symptoms which do not necessarily include pneumonia but can still lead to ICU hospitalization \citep{wadman2020does}.

\vspace{-5pt}
\subsection{COVID-19 Prediction from CXR}
\vspace{-5pt}
While many works have attempted to predict COVID-19 through medical imaging, the results have been on small or private data. Many of these methods use the dataset presented here.
For example, while promising results were achieved by \cite{Tarun2020Qure} (90\% AUC), these were reported over a large private dataset and are not reproducible. Additionally, much research is presented without appropriate evaluations, potentially leading to overfitting and performance overestimation \citep{Maguolo2020critic,Tartaglione2020Unveiling, DeGrave2020shortcut}. Many prediction models are not viable for use in clinical practice, as they are inadequately reported, particularly with regard to their performance, and at strong risk of bias \cite{Wynants2020criticalappraisal}. A study by \cite{Murphy2020} found that an AI system achieved an AUC of 0.81 and was comparable with that of six independent readers. 

Most ML work using this dataset has focused on COVID-19 Prediction similar to our task definition in \S \ref{sec:taskcovid}. Many have used a transfer learning approach, similar to what we use for baselines in this work, where a model that is pretrained on existing CXR datasets is used to construct features from the images in this dataset which are lower dimensional and are less prone to overfitting  \citep{Apostolopoulos2020covidtransferlearning}
\citep{Minaee2020covidtransferlearning}. Another similar approach taken is utilizing a multi-task network which predicts both standard CXR tasks as well as a COVID-19 task to assign a likelihood to samples and perform anomaly detection \citep{Zhang2020AnomalyDetection}. Another work has utilized the pathology hierarchy we provide to improve predictions by using the Clus-HMC method \citep{Pereira2020covidhierarchical}.

Many groups have also used this dataset to make severity and survival outcome predictions similar to our task definition in \S \ref{sec:taskseverity} and \ref{sec:tasksurvival}. Such work like \cite{Signoroni2020Brixia} focused on predicting a Brixia Score which has been clinically studied against severity and outcomes \citep{Borghesi2020BrixiaScore}. It is laborious to create these annotations so an approach by \cite{Amer2020severitylocalization} predicts a lung segmentation, generates a prediction saliency, and then calculates the ratio of coverage which acts as a proxy for the Brixia and geographic extent scores we discuss later.

\begin{figure}[t]
\begin{minipage}{0.60\textwidth}
\captionsetup{type=table}
\caption{Counts of each pneumonia frontal CXR by type and genus or species when applicable. The hierarchy structure is shown in the table from left to right. Information is collected by manually reading clinical notes for a mention of a confirmed test.}
\label{tab:stats}
\centering
\small
\begin{tabular}{llr}
\toprule
Type & Genus or Species & Image Count          \\
\midrule
          \cellcolor{orange!25}Viral & \cellcolor{orange!25}COVID-19 (SARSr-CoV-2) & \cellcolor{orange!25}      468 \\
          \cellcolor{orange!25}      & \cellcolor{orange!25}SARS (SARSr-CoV-1) &  \cellcolor{orange!25}      16 \\
          \cellcolor{orange!25}      & \cellcolor{orange!25}MERS-CoV &  \cellcolor{orange!25}      10 \\
          \cellcolor{orange!25}      & \cellcolor{orange!25}Varicella &  \cellcolor{orange!25}       5 \\
          \cellcolor{orange!25}      & \cellcolor{orange!25}Influenza & \cellcolor{orange!25}        4 \\
          \cellcolor{orange!25}      & \cellcolor{orange!25}Herpes & \cellcolor{orange!25}        3 \\
\cellcolor{red!25}Bacterial & \cellcolor{red!25}\emph{Streptococcus} spp. & \cellcolor{red!25}       13 \\
          \cellcolor{red!25}      & \cellcolor{red!25}\emph{Klebsiella} spp. & \cellcolor{red!25}        9 \\
          \cellcolor{red!25}      & \cellcolor{red!25}\emph{Escherichia coli} & \cellcolor{red!25}        4 \\
          \cellcolor{red!25}      & \cellcolor{red!25}\emph{Nocardia} spp. & \cellcolor{red!25}        4 \\
          \cellcolor{red!25}      & \cellcolor{red!25}\emph{Mycoplasma} spp. &  \cellcolor{red!25}       5 \\
          \cellcolor{red!25}      & \cellcolor{red!25}\emph{Legionella} spp. & \cellcolor{red!25}        7 \\
          \cellcolor{red!25}      & \cellcolor{red!25}Unknown & \cellcolor{red!25}        2 \\
          \cellcolor{red!25}      & \cellcolor{red!25}\emph{Chlamydophila} spp. & \cellcolor{red!25}        1 \\
          \cellcolor{red!25}      & \cellcolor{red!25}\emph{Staphylococcus} spp. & \cellcolor{red!25}        1 \\
\cellcolor{green!25}Fungal & \cellcolor{green!25}\emph{Pneumocystis} spp. & \cellcolor{green!25}       24 \\
            \cellcolor{green!25}  & \cellcolor{green!25}\emph{Aspergillosis} spp. & \cellcolor{green!25}       2 \\
\cellcolor{yellow!25}Lipoid & \cellcolor{yellow!25}Not applicable & \cellcolor{yellow!25}        8 \\
\cellcolor{yellow!25}Aspiration & \cellcolor{yellow!25}Not applicable & \cellcolor{yellow!25}        1 \\
\cellcolor{gray!25}Unknown     & \cellcolor{gray!25}Unknown & \cellcolor{gray!25}       59 \\
\bottomrule
\end{tabular}
\end{minipage}%
\begin{minipage}{0.4\textwidth}
\centering
\includegraphics[width=0.95\textwidth]{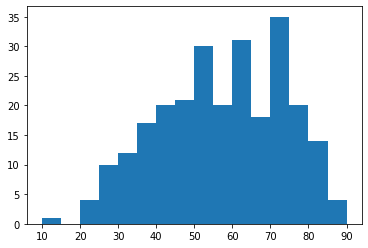}
\captionsetup{type=figure}
\vspace{-5pt}
\caption{Age per patient}
\label{fig:Histogram_Age}
\vspace{30pt}
\includegraphics[width=0.95\textwidth]{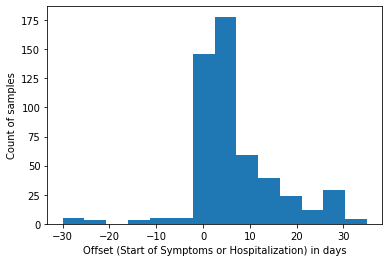}

\captionsetup{type=figure}
\vspace{-5pt}
\caption{Offset per image}
\label{fig:Histogram_Offset}
\end{minipage}
\vspace{-10pt}
\end{figure}

\vspace{-5pt}
\section{Cohort Details}
\vspace{-5pt}

The current statistics as of September 18th 2020 are shown in Table \ref{tab:stats}, which presents the distribution of frontal CXR by diagnosis, types of pneumonia and responsible microorganisms when applicable. For each image, attributes shown in Table \ref{tab:desc} are collected. Statistics are presented by sub-region in Table \ref{tab:cxr-region-stats} and by projection/view in Table \ref{tab:cxr-view-stats}.
Figures \ref{fig:Histogram_Age} and \ref{fig:Histogram_Offset} present demographics for patients and frontal CXR images. In terms of unique patients the M/F ratio is 230/139. In total, 761 images were collected, of which 679 are frontal and 82 are lateral view. Of the frontal views 518 are standard frontal PA/AP (Posteroanterior/Anteroposterior) views and 161 are AP Supine (Anteroposterior laying down). The images originate from various hospitals across 26 different countries.

\begin{table}[]
\caption{Descriptions of each attribute of the metadata}
\label{tab:desc}

\rowcolors{2}{gray!5}{white}
\centering
\resizebox{1\textwidth}{!}{%
\begin{tabular}{c p{400pt}}
\toprule
Attribute      & Description \\
\midrule
patientid& Internal identifier\\

offset& Number of days since the start of symptoms or hospitalization for each image. If a report indicates "after a few days", then 5 days is assumed.\\

sex& Male (M), Female (F), or blank\\

age& Age of the patient in years\\

finding &Type of pneumonia\\

RT\_PCR\_positive & Yes (Y) or no (N) or blank if not reported/taken\\

survival &If the patient survived the disease. Yes (Y) or no (N)\\

intubated&	Yes (Y) if the patient was intubated (or ventilated) at any point during this illness or No (N) or blank if unknown.\\
went\_icu&	Yes (Y) if the patient was in the ICU (intensive care unit) or CCU (critical care unit) at any point during this illness or No (N) or blank if unknown.\\
needed\_supplemental\_O2&	Yes (Y) if the patient required supplemental oxygen at any point during this illness or No (N) or blank if unknown.\\
extubated&	Yes (Y) if the patient was successfully extubated or No (N) or blank if unknown.\\
temperature	& Temperature of the patient in Celsius at the time of the image.\\

view&Posteroanterior (PA), Anteroposterior (AP), AP Supine (APS), or Lateral (L) for X-rays; Axial or Coronal for CT scans\\

modality& CT, X-ray, or something else\\

date& Date on which the image was acquired\\

location& Hospital name, city, state, country\\

filename& Name with extension\\

doi & Digital object identifier (DOI) of the research article\\

url & URL of the paper or website where the image came from\\

license& License of the image such as CC BY-NC-SA. Blank if unknown \\

clinical\_notes& Clinical notes about the image and/or the patient \\
other\_notes &e.g. credit \\

\bottomrule
\end{tabular}
}
\end{table}

\begin{table}[]
    \caption{Statistics with respect to sub-regions. Y/N counts represent only the subset of images which are labelled.}
    \label{tab:cxr-region-stats}
    \centering
\resizebox{1\textwidth}{!}{%
\begin{tabular}{llccccccccc}
\toprule
& &  & Frontal& & & Went & & & RT-PCR  &  \\
& & Patient & CXR  & Age   & Survival &  ICU & Intubation & COVID-19 & Positive & Male  \\
Region & Sub-region & Count   & Count  & Mean  & Y/N      & Y/N      & Y/N        & Count    & Count           & Ratio \\
\midrule
Africa                    & Northern Africa                 & 5       & 13     & 40.00    & 0/1      & 0/0      & 0/0        & 0        & 0               & 0.00     \\
\rowcolor{gray!5}Americas & Latin Amer./Caribbean & 5       & 5      & 42.80  & 1/0      & 1/1      & 0/1        & 3        & 1               & 0.80   \\
\rowcolor{gray!5} & Northern America                & 27      & 46     & 53.12 & 6/5      & 11/2\hphantom{0}     & 11/4\hphantom{0}       & 17       & 15              & 0.56  \\
Asia     & Eastern Asia                    & 54      & 90     & 55.14 & 16/6\hphantom{0}     & 6/5      & 8/6        & 50       & 40              & 0.35  \\
                          & South-eastern Asia              & 9       & 17     & 46.67 & 4/1      & 1/3      & 1/3        & 8        & 8               & 0.56  \\
                          & Southern Asia                   & 15      & 24     & 48.27 & 4/0      & 3/0      & 3/0        & 14       & 14              & 0.73  \\
                          & Western Asia                    & 13      & 20     & 46.92 & 2/0      & 3/0      & 2/0        & 6        & 4               & 0.54  \\
\rowcolor{gray!5} Europe   & Eastern Europe                  & 4       & 7      & 67.50  & 1/0      & 0/0      & 0/0        & 1        & 1               & 0.50   \\
\rowcolor{gray!5}  & Northern Europe                 & 31      & 60     & 58.78 & 3/3      & 7/0      & 7/1        & 22       & 18              & 0.55  \\
\rowcolor{gray!5}  & Southern Europe                 & 123     & 195    & 59.30  & 31/4\hphantom{0}     & 25/9\hphantom{0}     & 13/22      & 91       & 74              & 0.63  \\
\rowcolor{gray!5} & Western Europe                  & 53      & 92     & 52.83 & 19/2\hphantom{0}     & 19/29    & 1/1        & 49       & 5               & 0.60   \\
Oceania                   & Australia/New Zealand       & 42      & 60     & 50.48 & 4/0      & 1/1      & 1/1        & 1        & 2               & 0.57  \\
\rowcolor{gray!5}Unknown                   & Unknown                         & 31      & 50     & 53.64 & 3/3      & 6/0      & 7/2        & 19       & 0               & 0.55  \\
\midrule
Totals & & 412 & 679 & 55.1 & 119 & 133 & 95 & 281 & 182 & 0.55 \\
\bottomrule
\end{tabular}
}
\end{table}

\begin{table}[ht]
\caption{Statistics with respect to the projection/view of the CXR. Y/N counts represent only the subset of images which are labelled.}
\label{tab:cxr-view-stats}
\centering
\resizebox{1\textwidth}{!}{%
\begin{tabular}{lcccccccc}
\toprule
{} & Patient &    Age & Survival  & ICU  & Intubation  & COVID-19 & RT-PCR  & Male  \\
View/Projection  &    Count &   Mean &           Y/N &       Y/N &              Y/N &      Count &             Positive &       Ratio \\
\midrule
PosteroAnterior (PA)  & 326      & 52.21 & 104/19\hphantom{0}        & 53/67        & 33/55          & 191      & 112             & 0.54     \\
AnteroPosterior (AP)      & 192      & 57.93 & 36/13        & 58/6\hphantom{0}          & 51/15          & 141      & 109             & 0.54       \\  
AP Supine & 161      & 57.84 & 51/20        & 89/11        & 42/6\hphantom{0}          & 136      & 64              & 0.64       \\
Lateral (L)      & 82       & 49.78 & 25/2\hphantom{0}          & \hphantom{0}5/11         & \hphantom{0}2/16           & 25       & 20              & 0.73       \\
\bottomrule
\end{tabular}
}
\end{table}

\subsection{Data Collection}
As mentioned earlier, data was largely compiled from public databases on websites such as Radiopaedia.org, the Italian Society of Medical and Interventional Radiology\footnote{https://www.sirm.org/category/senza-categoria/covid-19/}, Figure1.com\footnote{https://www.figure1.com/covid-19-clinical-cases}, and the Hannover Medical School \citep{Winther2020ovid-19-image-repository}, both manually and using scrapers. A full list of publications is included in Appendix \S \ref{sec:sourcepapers}.

Images were extracted from online publications, websites, or directly from the PDF using the tool pdfimages\footnote{https://poppler.freedesktop.org/}. Throughout data collection, we aimed to maintain the quality of the images. Many articles were found using the list of literature provided by \cite{Peng2020COVID-19-CT-CXR}. A limitation of this approach is that we have no control over the processing between the PAC system and the PDF or website. However, we believe that information about the radiological finding was maintained otherwise the image in publication would not be useful. We also believe the images from PDFs and other image sources provide useful information when compared to raw data and expect studies to show this.

A challenge in extracting metadata is the alignment with images. To belong in the same row as an image, a clinical measurement must have been taken on the same day. It can be difficult to automatically determine when the measurement was taken if the metadata appears outside of image captions. For details on scraper design, see \S Appendix \ref{apdx:scraper}. All scripts are made publicly available.

\section{Experimental Setup}
Using this dataset as a benchmark is very challenging because, by the nature of its construction, it is very biased and unbalanced (recall geographically unbalanced labels in Table \ref{tab:cxr-region-stats}). This can lead to many negative outcomes if treated as a typical benchmark dataset \citep{Maguolo2020critic, cohen2020limits, Seyyed-Kalantari2020CheXclusion,Kelly2019challenges}. Unless otherwise specified only AP and PA views are used to avoid confounding image artifacts of the AP Supine view.

\subsection{Leave-One-Country/Continent-Out (LOCO) Evaluation:} 
In order to deal with issues of bias, we perform a ``leave one country/continent out'' (LOCO) evaluation. This approach is motivated from training bias common in small, unbalanced datasets. For our evaluation the test set will be composed of data from a single continent. Ideally, we would separate by countries, but this is not possible given the current distribution of data. We also note that not every sample is labelled, so models may be trained and evaluated on data from the same continent, but we ensured that samples in training and evaluation did not originate from the same research group/image source.\\ 
This approach should give us a distributional shift that will allow us to correctly evaluate the model. For each task, a subset of the samples will have enough representation to be included. Continents which do not have at least one representative for each class are filtered out and not used.

\subsection{Models and Features}
In place of images, features are extracted using a pre-trained DenseNet model \citep{Huang2017densenet} from the TorchXRayVison library \citep{Cohen2020xrv} which is trained on 7 large CXR datasets as described in \cite{cohen2020limits}. These datasets were manually aligned with each other on 18 common radiological finding tasks in order to train a model using all datasets at once (atelectasis, consolidation, infiltration, pneumothorax, edema, emphysema, fibrosis, fibrosis, effusion, pneumonia, pleural thickening, cardiomegaly, nodule, mass, hernia, lung lesion, fracture, lung opacity, and enlarged cardiomediastinum). For example ``pleural effusion" from one dataset is considered the same as ``effusion" from another dataset in order to consider these labels equal. In total, 88,079 non-COVID-19 images were used to train the model on these tasks. 
We will use the term pre-sigmoid which refers to the output of the last layer of the network multiplied by a weight vector corresponding to each task which would normally then be passed through a sigmoid function.
Features will be used in the following constructions (as in \cite{Cohen2020severity}): 

\begin{itemize}
\item \textbf{Intermediate features} - the result of the convolutional layers and global averaging (1024 dim vector);
\item \textbf{18 outputs} - each image was represented by the 18 outputs (pre-sigmoid) here: Atelectasis, Consolidation, Infiltration, Pneumothorax, Edema, Emphysema, Fibrosis, Effusion, Pneumonia, Pleural Thickening, Cardiomegaly, Nodule, Mass, Hernia, Lung Lesion, Fracture, Lung Opacity, Enlarged Cardiomediastinum;
\item \textbf{4 outputs} - a hand picked subset of the above mentioned outputs (pre-sigmoid) were used containing radiological findings more frequent in pneumonia: Lung Opacity, Pneumonia, Infiltration, and Consolidation;
\item \textbf{Lung Opacity output} - the single output (pre-sigmoid) for lung opacity was used because it was very related to this task. This feature was different from the opacity score that we would like to predict;
\item \textbf{Image pixels} - the image itself as a vector of pixels (224$\times$224=50176).
\item \textbf{Baseline prevalence} - to compare against a model which predicted based on the prevalence of the labels only.
\end{itemize}

Model training is done using linear or logistic regression with default parameters from Sci-kit learn \citep{scikit-learn}. An L2 penalty was used with an LBFGS solver. In the case of ``image pixels'' an MLP is used with 100 hidden units and  ReLU activations \cite{Glorot2011ReLU}. It is trained with full batch gradient descent using an Adam optimizer \cite{Kingma2014adam} (with $\beta_1=0.9, \beta_2=0.999, \epsilon=1e^{-8}$, a learning rate of $0.001$, and 10\% of the data used for early stopping.

\section{Task Ideas with Baseline Evaluations}

We present multiple clinical use cases and potential tools which could be built using this dataset and present a baseline task which is evaluated. We describe the scenarios in detail to both convey what our group has learned while interacting with clinicians as well as solidify what the value is of such a model to avoid misguided efforts solving a problem that doesn't exist. For the results presented in Tables \ref{tab:regression} and \ref{tab:classification} a full listing over all data splits is presented in Appendix \S \ref{sec:fullresults}.

\begin{table}[h]
\centering
\small
\vspace{-15pt}
\caption{Classification tasks. Evaluation is performed using LOCO evaluation. The metrics here are the average over each hold out countries test set. AUROC is the area under the TPR-FRP (ROC) curve, and  AUPRC is the area under the precision-recall curve. The scores are averages over the held out test countries.}
\label{tab:classification}
\begin{tabular}{cccccc}
\toprule
Task& Test Regions& Features & \# params & AUROC &      AUPRC  \\

\midrule
\multirow{6}{*}{\begin{tabular}[]{@{}c@{}}
COVID-19\\0<offset<8 days\\True=128\\False=27
\end{tabular}}
& \multirow{6}{55pt}{Americas, Asia, Europe, Oceania}
  & 18 outputs & 18+1 &  0.58$\pm$0.12 &  0.78$\pm$0.16 \\
  &   & Intermediate features & 1024+1 &  0.57$\pm$0.12 &  0.77$\pm$0.15 \\
  &   & 4 outputs & 4+1 &  0.50$\pm$0.01 &  0.73$\pm$0.22 \\
  &   & Baseline prevalence & 1+1 &  0.50$\pm$0.00 &  0.73$\pm$0.22 \\
  &   & lung opacity output & 1+1 &  0.50$\pm$0.01 &  0.73$\pm$0.22 \\
  &   & Image pixels (MLP) & 5017801 &  0.48$\pm$0.06 &  0.72$\pm$0.20 \\

\midrule
\multirow{6}{*}{\begin{tabular}[]{@{}c@{}}
Viral or Bacterial\\0<offset<8 days\\True=140\\False=10
\end{tabular}}
& \multirow{6}{55pt}{Europe, Oceania}
  & 18 outputs & 18+1 &  0.82$\pm$0.26 &  0.97$\pm$0.04 \\
  &   & Intermediate features & 1024+1 &  0.70$\pm$0.19 &  0.84$\pm$0.13 \\
  &   & lung opacity output & 1+1 &  0.56$\pm$0.08 &  0.72$\pm$0.31 \\
  &   & 4 outputs & 4+1 &  0.52$\pm$0.03 &  0.71$\pm$0.30 \\
  &   & Baseline prevalence & 1+1 &  0.50$\pm$0.00 &  0.71$\pm$0.30 \\
  &   & Image pixels (MLP) & 5017801 &  0.50$\pm$0.01 &  0.71$\pm$0.23 \\

\midrule
\multirow{6}{*}{\begin{tabular}[c]{@{}c@{}}
Survival prediction\\0<offset<8 days\\True=46\\False=7
\end{tabular}}
& \multirow{6}{55pt}{Americas, Asia, Europe}
  & lung opacity output & 1+1 &  0.55$\pm$0.08 &  0.86$\pm$0.05 \\
  &   & 4 outputs & 4+1 &  0.54$\pm$0.07 &  0.85$\pm$0.05 \\
  &   & Baseline prevalence & 1+1 &  0.50$\pm$0.00 &  0.84$\pm$0.04 \\
  &   & Image pixels (MLP) & 5017801 &  0.50$\pm$0.00 &  0.84$\pm$0.03 \\
  &   & Intermediate features & 1024+1 &  0.50$\pm$0.00 &  0.84$\pm$0.04 \\
  &   & 18 outputs & 18+1 &  0.47$\pm$0.03 &  0.84$\pm$0.03 \\

\midrule
\multirow{6}{*}{\begin{tabular}[c]{@{}c@{}}
ICU stay\\0<offset<8 days\\True=17\\False=27
\end{tabular}}
& \multirow{6}{55pt}{Asia, Europe}
  & 18 outputs & 18+1 &  0.81$\pm$0.10 &  0.50$\pm$0.24 \\
  &   & Intermediate features & 1024+1 &  0.71$\pm$0.18 &  0.40$\pm$0.21 \\
  &   & 4 outputs & 4+1 &  0.55$\pm$0.15 &  0.35$\pm$0.36 \\
  &   & Baseline prevalence & 1+1 &  0.50$\pm$0.00 &  0.28$\pm$0.26 \\
  &   & lung opacity output & 1+1 &  0.42$\pm$0.12 &  0.28$\pm$0.26 \\
  &   & Image pixels (MLP) & 5017801 &  0.40$\pm$0.13 &  0.28$\pm$0.20 \\

\midrule
\multirow{6}{*}{\begin{tabular}[c]{@{}c@{}}
Intubated\\0<offset<8 days\\True=12\\False=23
\end{tabular}}
& \multirow{6}{55pt}{Asia, Europe}
  & Intermediate features & 1024+1 &  0.61$\pm$0.06 &  0.40$\pm$0.21 \\
  &   & Baseline prevalence & 1+1 &  0.50$\pm$0.00 &  0.32$\pm$0.21 \\
  &   & Image pixels (MLP) & 5017801 &  0.50$\pm$0.07 &  0.32$\pm$0.17 \\
  &   & lung opacity output & 1+1 &  0.50$\pm$0.00 &  0.32$\pm$0.21 \\
  &   & 18 outputs & 18+1 &  0.45$\pm$0.22 &  0.35$\pm$0.26 \\
  &   & 4 outputs & 4+1 &  0.45$\pm$0.00 &  0.31$\pm$0.20 \\
\bottomrule
\end{tabular}
\end{table}

\subsection{Complement to COVID-19 Pneumonia Diagnosis and Management}
\label{sec:taskcovid}

\paragraph{Motivation:}%
While reverse transcriptase polymerase chain reaction (RTPCR) assay remains the gold standard for diagnosis, CXR plays a major role as the top initial imaging test for patients with suspected COVID-19 pneumonia \citep{Dennie2020-CSTR-CAR-Statement}. Because of their relative lack of sensitivity (69\%, although this was computed with a small sample size) \citep{10.1148/radiol.2020201160} and the fact that they are often normal early in the disease \citep{Dennie2020-CSTR-CAR-Statement,10.1148/radiol.2020201160}, a negative CXR should not be used to rule out COVID-19 infection \citep{Dennie2020-CSTR-CAR-Statement}. Instead, features of COVID-19 pneumonia in CXR, although nonspecific, raise the pretest probability of infection. For example, distinguishing between viral and bacterial pneumonia could influence management in addition to other clinical clues \citep{Heneghan2020viralfrombacterialCEBM}. According to the CSTR and CAR, a CXR is ``most useful when an alternative diagnosis is found that completely explains the patient’s presenting symptoms such as, but not limited, to pneumothorax, pulmonary edema, large pleural effusions, lung mass or lung collapse \citep{Dennie2020-CSTR-CAR-Statement}.''

\paragraph{Task Specification:} The hierarchy of labels (Table \ref{tab:stats}) allows us to perform multiple different classification tasks. A first task is to classify COVID-19 from other causal agents of pneumonia such as bacteria or other viruses.
A second task is to distinguish viral from bacterial pneumonia.

\paragraph{Results:} When classifying between COVID/non-COVID most recent works using this dataset have taken other datasets and treated them as non-COVID-19 \citep{Wang2020covidnet,Apostolopoulos2020covidtransferlearning} while results with balanced datasets from \cite{Tartaglione2020Unveiling} report much lower performance. Our LOCO evaluation aims to avoid these issues and in Table \ref{tab:classification} we find that the best performance is only slightly higher than random guessing which would yield a 0.5 AUROC. We specifically note that the ``4 outputs'' which are commonly associated with Pneumonia are not predictive, possibly implying something about the disease that should be explored more. We find that predicting between bacterial and viral yields reasonably good performance using intermediate features or the 18 specific outputs. This is not surprising as previous work has reported high accuracy when predicting bacterial and viral from pediatric CXR \citep{Kermany2018}.

\subsection{Severity Prediction, Including Intensive Care Unit (ICU) Stay}
\label{sec:taskseverity}
\paragraph{Motivation:}
The ICU is reserved for patients who require life support such as mechanical ventilation. In this invasive intervention reserved for patients unable to breathe on their own, an endotracheal tube is inserted into the windpipe (intubation) and the lungs are mechanically inflated and deflated \citep{Tobin2017MechanicalVentilation}. Predicting the need for mechanical ventilation in advance could help plan management or prepare the patient. Another challenge is knowing when to remove mechanical ventilation (extubation), which falls in a specific window of time \citep{Thille2013Extubate}.

Assessing the severity of a patient's condition using a well adopted scoring system is a key aspect of patient management \citep{10.1148/radiol.2020201160,Cohen2020severity, Borghesi2020BrixiaScore,Signoroni2020Brixia}. A model which predicts the severity of COVID-19 pneumonia, and pneumonia in general, based on CXR could be used as an assistive tool when managing patient care for escalation or de-escalation of care, especially in the ICU.

Non-ML work by \cite{Allenbach2020ICUTransfer} combines information from CT scans and CXR with other clinical information to create a score-based predictive model for transfer to the ICU. 

This task is further motivated by papers published after a preprint of this work which employed siamese networks to represent severity \citep{Li2020SiameseCXR} as well as predicted deterioration risk scores and time estimates \citep{Shamout2020deterioration}.

It is important to note with this task that predicting these events could be confounded by the presence of an endotracheal tube in the CXR of a mechanically ventilated patient; when available, this is annotated in our dataset.

Due to the reduced mobility of ICU patients, CXR are often obtained with the patient lying down in a view referred to as ``AP supine'' \citep{Khan2009ReadingRadiographs}, which includes but is not limited to patients who are intubated or soon to be. Because this position drastically modifies the appearance of the CXR, a naive approach could confound such changes with the need to be intubated.

\paragraph{Task Specification:}
The first thing that is required for this task is a well defined scoring system that defines patient severity. The mRALE (modified Radiographic Assessment of Lung Edema) score  \citep{10.1148/radiol.2020201160,Cohen2020severity} was developed specifically with in context of assessing COVID-19 severity from CXR and was used to score 94 images in this dataset in \cite{Cohen2020severity} by two chest radiologists and a radiology resident. Also, the Brixia score  \citep{Borghesi2020BrixiaScore,Signoroni2020Brixia} was used by another group to score 192 of these images by "two expert radiologists, a board-certified staff member and a trainee with 22 and 2 years of experience respectively" \citep{Signoroni2020Brixia}.

Our task and evaluation will use the severity scores created by \cite{Cohen2020severity}. This work provides two scores: Geographic Extent (0-8), how much opacity covers the lungs, and Opacity (0-6), how opaque the lungs are.

Also, an ICU stay and the patient being intubated are both predicted given patients between 0 and 8 days from symptoms/presentation. Images marked as intubation present are excluded from the evaluation as this would be a visible confounder in the image. For ICU stay prediction images marked as already in the ICU are excluded.

\paragraph{Results:}
Table \ref{tab:regression} shows performance similar to that reported by \cite{Cohen2020severity}. The evaluation used in that work is not LOCO.

Table \ref{tab:classification} shows ICU stay is predicted reasonably well at 0.81 AUROC using all 18 outputs. The improvement over the ``4 outputs'' which are associated with Pneumonia may imply that ICU stay is predictive by features not related to  pneumonia. Intubation is predicted poorly.

\begin{table}[t]
\centering
\vspace{-15pt}
\caption{Regression Tasks. Evaluation is performed using LOCO evaluation. The metrics here are the average over each held-out-continents test set. $R^2$ : coefficient of determination; MAE: mean absolute error. ``4 outputs" refers to Lung Opacity, Pneumonia, Infiltration, and Consolidation.}
\label{tab:regression}
\vspace{-5pt}
\resizebox{1\textwidth}{!}{%
\begin{tabular}{ccccccc}
\toprule
Task & Test Regions & Features & \begin{tabular}[c]{@{}c@{}}\# of \\ parameters\end{tabular} & \begin{tabular}[c]{@{}c@{}}Pearson\\Correlation\end{tabular} & $R^2$ & MAE \\

\midrule
\multirow{5}{*}{\begin{tabular}[c]{@{}c@{}}Geographic\\Extent\\Score (0-8)\\N=94\end{tabular}} 
& \multirow{5}{65pt}{Asia, Europe}
  & 4 outputs & 4+1 &  0.82$\pm$0.05 &   0.62$\pm$0.05 &  1.11$\pm$0.08 \\
  &   & 18 outputs & 18+1 &  0.80$\pm$0.09 &   0.59$\pm$0.17 &  1.15$\pm$0.15 \\
  &   & lung opacity output & 1+1 &  0.80$\pm$0.05 &   0.58$\pm$0.02 &  1.15$\pm$0.01 \\
  &   & Intermediate features & 1024+1 &  0.77$\pm$0.06 &   0.41$\pm$0.17 &  1.41$\pm$0.08 \\
  &   & Baseline prevalence & 1+1 &  0.00$\pm$0.00 &  -0.33$\pm$0.25 &  2.14$\pm$0.31 \\

\midrule
\multirow{5}{*}{\begin{tabular}[c]{@{}c@{}}Opacity\\Score (0-6)\\N=94\end{tabular}}
& \multirow{5}{65pt}{Asia, Europe}
 & 4 outputs & 4+1 &  0.79$\pm$0.07 &   0.61$\pm$0.11 &  0.73$\pm$0.10 \\
  &   & lung opacity output & 1+1 &  0.79$\pm$0.07 &   0.60$\pm$0.09 &  0.76$\pm$0.10 \\
  &   & 18 outputs & 18+1 &  0.66$\pm$0.13 &   0.29$\pm$0.26 &  0.90$\pm$0.16 \\
  &   & Intermediate features & 1024+1 &  0.68$\pm$0.11 &  -0.09$\pm$0.45 &  1.20$\pm$0.26 \\
  &   & Baseline prevalence & 1+1 &  0.00$\pm$0.00 &  -0.26$\pm$0.20 &  1.30$\pm$0.00 \\

\bottomrule
\end{tabular}
}
\vspace{-5pt}
\end{table}

\subsection{Survival outcome}
\label{sec:tasksurvival}

\paragraph{Motivation:}
Not too dissimilar to severity prediction, determining at what point survival can be predicted could be useful for patient management. Given a series of patient chest X-rays over time, it could be possible to determine the probability of survival.  

Non-ML models are able to predict in-hospital mortality for patients with COVID-19 using an original severity score for CXR (Brixia score) combined with two predictive factors, which are patient age and immunosuppressive conditions  \citep{10.1016/j.ijid.2020.05.021}.
ML models are able to predict survival based on clinical features (lactate dehydrogenase, lymphocyte count, and high-sensitivity C-reactive protein) with high accuracy \citep{Yan2020PredictionCriticality}.

\paragraph{Task Specification:}
In order to make predictions which are relevant to the clinical context, it is important to control for the time period when the patient is observed. Our evaluation is on data between 0 and 8 days since symptoms or admission in order to simulate predicting at the beginning of management. Predictions will be made on non-intubated patients to avoid this confounder of severity.

\paragraph{Results:}
In Table \ref{tab:classification}, almost random performance of 0.55 AUROC is obtained using lung opacity and the 4 hand picked features of Pneumonia. In general, no method works well.

\subsection{Trajectory Prediction}

\paragraph{Motivation:}
The ability to gauge severity of COVID-19 lung infections could be used to complement other severity tools for escalation or de-escalation of care, especially in the ICU.
Following diagnosis, patients' CXR could be scored periodically to objectively and quantitatively track pulmonary disease progression and treatment response. Eventually, physicians could track patients' response to various drugs and treatments using CXR and uploading the images to the dataset, allowing researchers to create predictive tools to measure recuperation. Such a model could also be used as an objective tool to compare response to different management algorithms and 
inspire better management strategies.

\begin{figure}[t]
\vspace{-20pt}
\centering
    \subfigure[Illustration of Idea]{
    \includegraphics[width=0.475\textwidth]{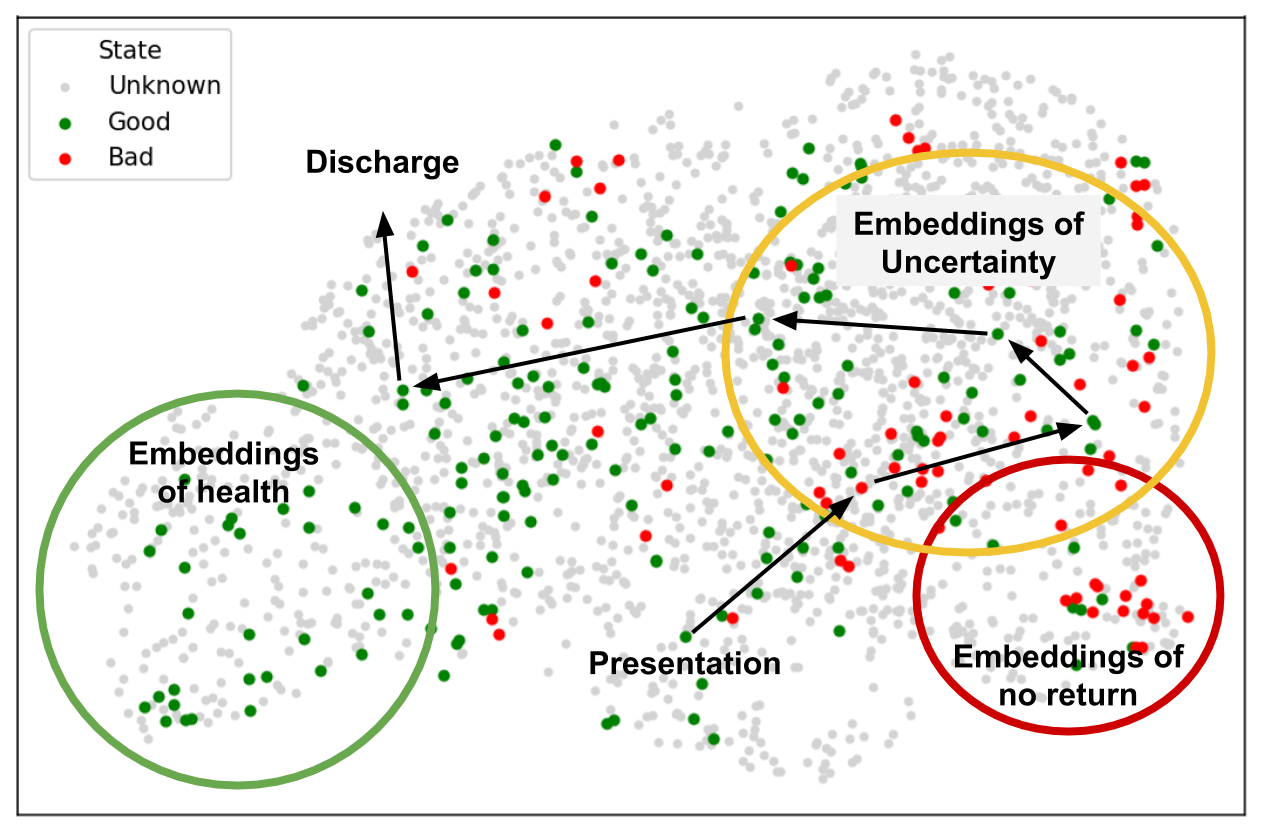}
    }%
    \subfigure[Current realization]{
    \includegraphics[width=0.495\textwidth]{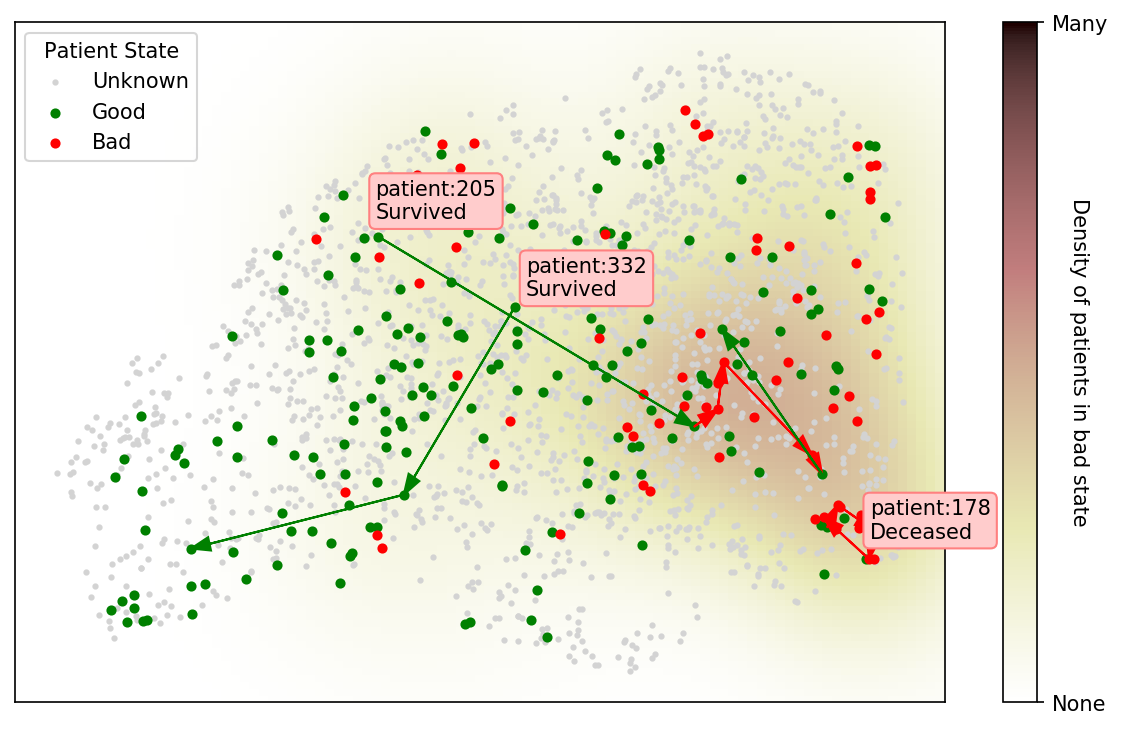}
    }%
    \caption{A UMAP \citep{McInnes2018UMAP} visualization of each CXR from this dataset together with the Kaggle RSNA pneumonia images. CXRs with a trajectory are shown with an arrow between timepoints. If survival outcome is known the arrows and/or points are colored. The background is colored based on the density of points in the ICU or that are intubated.}
    \label{fig:trajectory}
\end{figure}

If the representation is expressive enough, patients can be plotted as shown in Figure \ref{fig:trajectory}a. A conceptual figure and our current realization are shown using a pre-trained CXR model showing the available trajectories and patient outcomes. This approach could serve as a way to iterate quickly with a medical team (to simply explore the learned representation instead of building complete tools) to make sense of the complexities of these models and patients. 

\paragraph{Task Specification:}
Using this dataset, we could visualize a model's representation of CXRs and plot the trajectory of patients according to a color scheme representing their state/image (good or bad). The representation is based off of the 18 pre-sigmoid outputs of the DenseNet discussed. A good state is defined as non-intubated, not in the ICU, or the last state before discharge. A bad state is defined as intubated, in the ICU, or any state which is the last in the series and the patient is determined to have died. A kernel density estimation is taken of the 2d embeddings of all bad states to illustrate severity. Here PA, AP, and AP Supine images are used to maximize the data visualized as we did not observe any shift in the representation.

\paragraph{Results:}
Figure \ref{fig:trajectory}b shows proximity of image embeddings which represent patients in a good or bad state, demonstrating the potential insight already contained in these pre-trained models. The three patients display trajectories as we would imagine. Patient 178\footnote{\href{https://www.eurorad.org/case/16660
}{https://www.eurorad.org/case/16660}} spirals in an area which appears bad. Patient 205\footnote{\href{https://radiopaedia.org/cases/covid-19-pneumonia-progression-and-regression
}{https://radiopaedia.org/cases/covid-19-pneumonia-progression-and-regression
}} progresses into bad states but manages to pull themselves out. Patient 332 \citep{10.4269/ajtmh.20-0203} seems to recover quickly to a region which is only populated by good states.

\section{Future Task Ideas}
There are many potential tasks that we were not able to explore in this work, primarily the use of saliency maps and the utility of out-of-distribution models.

Evaluating saliency maps can be challenging, but it is a useful evaluation for methods which aim to explain model predictions \citep{Taghanaki2019InfoMask, Singla2020Exaggeration}. Our dataset contains lung bounding boxes, which were contributed by Andrew Gough at General Blockchain Inc. for 167 images annotated for the left and right lung. Also, 209 generated lung segmentations were added to the dataset by \cite{Selvan2020lungVAE}; the team trained a model on an external dataset and applied it to our dataset. As there are no ``ground truth'' segmentations, these can be used to examine if saliency maps are located reasonably within lung regions to detect overfitting. This can be seen by checking if the predictive regions of the image lie outside of the region of interest \citep{Ross2017rrr, Badgeley2019, Viviano2019}.

General anomaly detection could be a useful model when trying to identify what is new about an illness such as COVID-19. Out-of-distribution (OoD) tools \citep{Shafaei2018} or unpaired distribution matching models \citep{Zhu2017cyclegan, Kim2017discogan} could capture the shift in distributions and present them as changes to images \cite{cohen2018distribution}. Identifying what about the COVID-19 distribution is different from other viral or bacterial pneumonias could aid in studying both the disease as well as the models representations for overfitting \citep{Singla2020Exaggeration}. Transfer learning methods actively under development in the ML community such as few/zero-shot \citep{Wang2019fewshot,Tian2020fewshot,Larochelle2008zerodata,Ren2019IncrementalFew-Shot}, meta-learning \citep{Andrychowicz2016,Snell2017Prototypical}, deep metric learning \citep{Roth2020dml}, and domain adaptation \citep{Motiian2017Few-ShotDA} will likely be useful in this setting. 

\section{Conclusion}
\vspace{-5pt}

This paper presents a dataset of COVID-19 images together with clinical metadata which can be used for a variety of tasks. This dataset puts existing ML algorithms to the test. Given the number of existing large CXR datasets, novel tasks related to COVID-19 present a relevant challenge to overcome. We note that a major limitation of this work is the selection bias when gathering publicly available images which are likely made public for educational reasons because they are clear examples or interesting cases. Therefore, they do not represent the real world distribution of cases. Furthermore, another selection bias is that the information given on public platforms such as Figure 1 or Radiopaedia might not be complete for all patients and/or omit normal values (e.g., presence or absence of transfer to the ICU, lymphocyte count). Lastly, we do not yet have variables such as ethnic background, pre-existing conditions, and immunosuppression status. Any clinical claims made from models must therefore be backed by rigorous evaluation and take into account these limitations. Nevertheless, we believe that this dataset and the discussion of clinical context will contribute towards the machine learning community developing solutions with potential use in healthcare.

\section*{Broader Impact}
\vspace{-5pt}

This project aims to make a dataset of patients with a novel life-threatening disease accessible to researchers so that tools can be created to aid in the care of future patients. The manner in which we collect existing public data ensures that patients are not put at risk. 

Data impact: Image data linked with clinically relevant attributes in a public dataset that is designed for ML will enable parallel development of diagnosis and management tools and rapid local validation of models. Furthermore, this data can be used for a variety of different tasks.

Tool impact: Tools developed using this data and with the ideas presented can give physicians an edge and allow them to act with more confidence while they wait for the analysis of a radiologist by having a digital second opinion confirm their assessment of a patient's condition. In addition, these tools can provide quantitative scores which can enable large scale analysis of CXR without the need for costly/time consuming manual annotations.

\section*{Acknowledgements}
We thank Dr. Errol Colak, Luke Oakden-Rayner, Rupert Brooks, Hadrien Bertrand, Dr. Michaël Chassé, and Dr. Carl Chartrand-Lefebvre for their input. This research is based on work partially supported by the CIFAR AI and COVID-19 Catalyst Grants. This work utilized the supercomputing facilities managed by Compute Canada and Calcul Quebec. We thank AcademicTorrents.com for making data available for our research. 

%
\ethics{The work follows appropriate ethical standards in conducting research and writing the manuscript, following all applicable laws and regulations regarding treatment of animals or human subjects. This project is approved by the University of Montreal's Ethics Committee \#CERSES-20-058-D}

\coi{We declare we don't have conflicts of interest.}

\small
\bibliography{refs,chester-covid-19,papers, papers2, newreferences}

\newpage
\appendix
\section*{Appendix}

\section{Extra dataset statistics}
\label{sec:stats}

Just considering PA, AP, and AP Supine views there are 367 JPEG files and 171 PNGs. All images are 8-bit except for 1 which is 16-bit.

\begin{figure}[h]
\centering
    \subfigure[Heights]{
    \includegraphics[width=0.48\columnwidth]{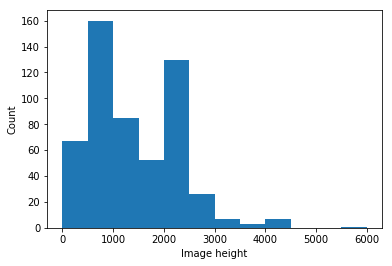}
    }%
    \subfigure[Widths]{
    \includegraphics[width=0.48\columnwidth]{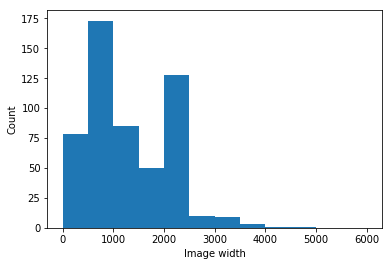}
    }%
    
    \caption{Histograms of image sizes in pixels.}
    \label{fig:imagesizes}
\end{figure}

\newpage
\section{Example images}
\label{sec:examples}

\begin{figure}[h]
\centering
    \subfigure[Day 10]{
    \includegraphics[width=0.235\columnwidth]{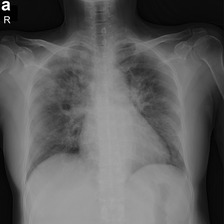}
    }%
    \subfigure[Day 13]{
    \includegraphics[width=0.235\columnwidth]{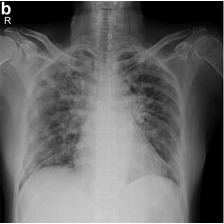}
    }%
    \subfigure[Day 17]{
    \includegraphics[width=0.235\columnwidth]{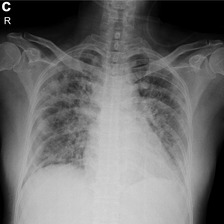}
    }%
    \subfigure[Day 25]{
    \includegraphics[width=0.235\columnwidth]{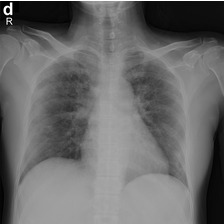}
    }%
    
    \caption{Example images from the same patient (\#19) extracted from \citet{10.1016/j.jfma.2020.02.007}. This 55 year old female survived a COVID-19 infection.}
    \label{fig:example}
\end{figure}

\begin{figure}[h]
\centering
    \subfigure[Day 1]{\includegraphics[width=0.20\columnwidth]{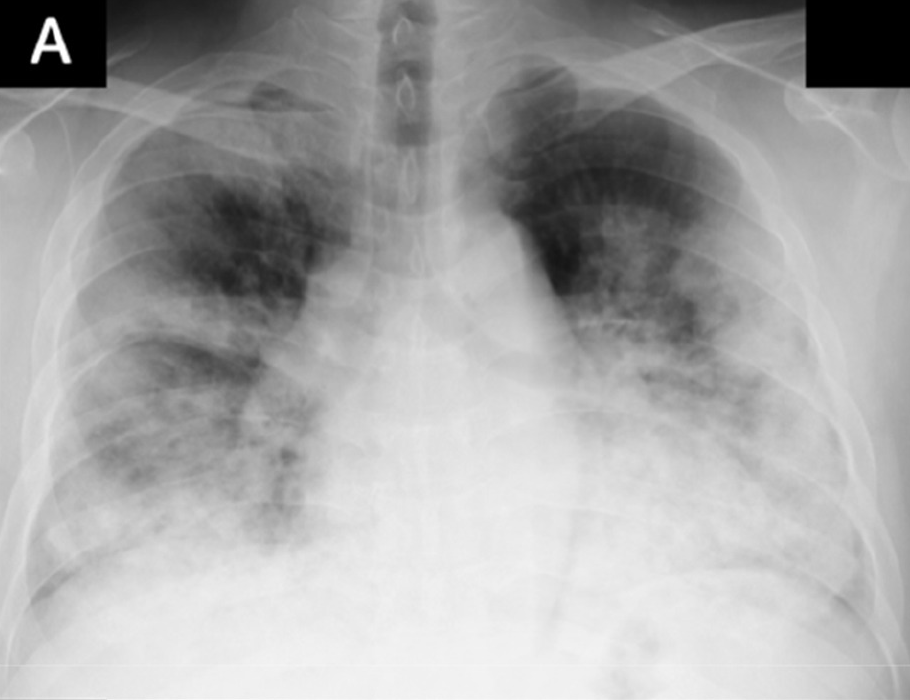}}%
    \subfigure[Day 2]{\includegraphics[width=0.20\columnwidth]{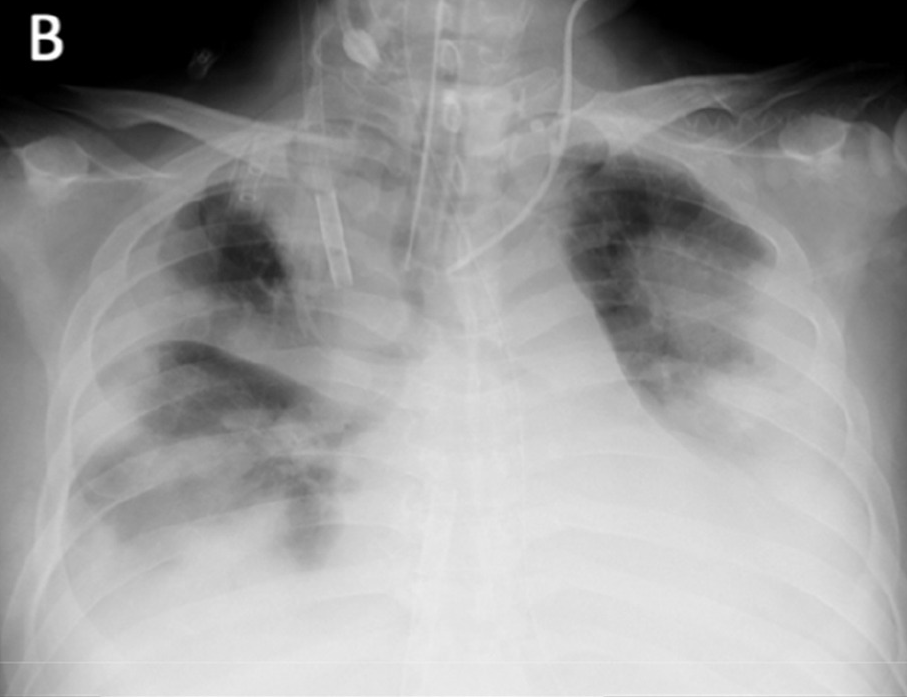}}%
    \subfigure[Day 7]{\includegraphics[width=0.20\columnwidth]{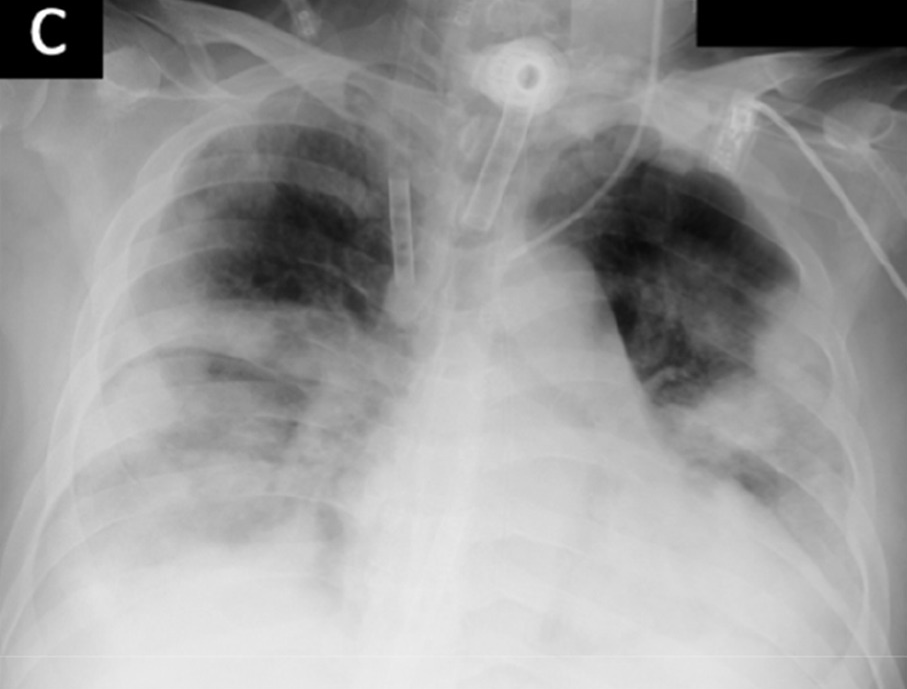}}%
    \subfigure[Day 12]{\includegraphics[width=0.20\columnwidth]{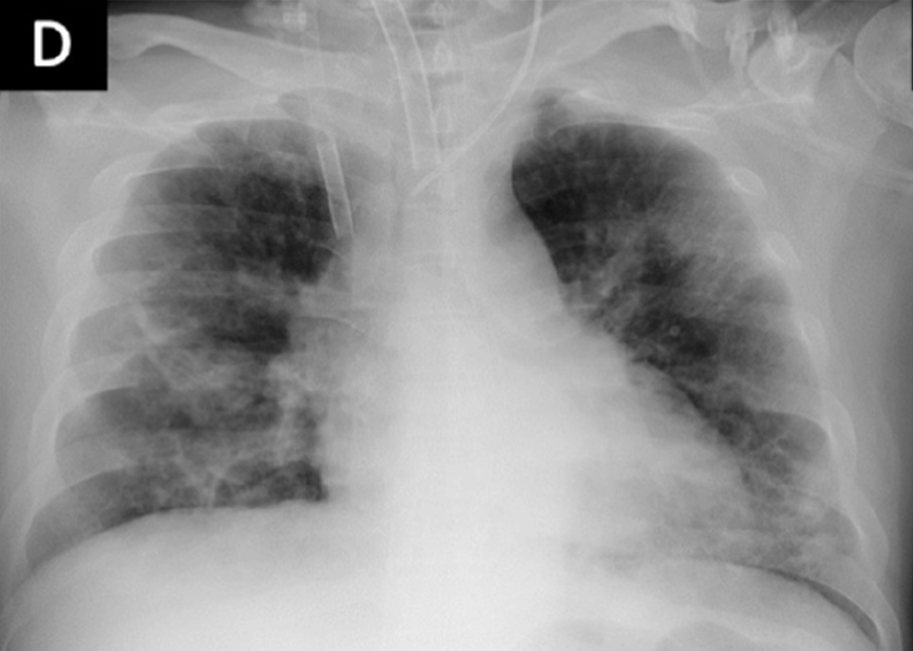}}%
    \subfigure[Day 17]{\includegraphics[width=0.20\columnwidth]{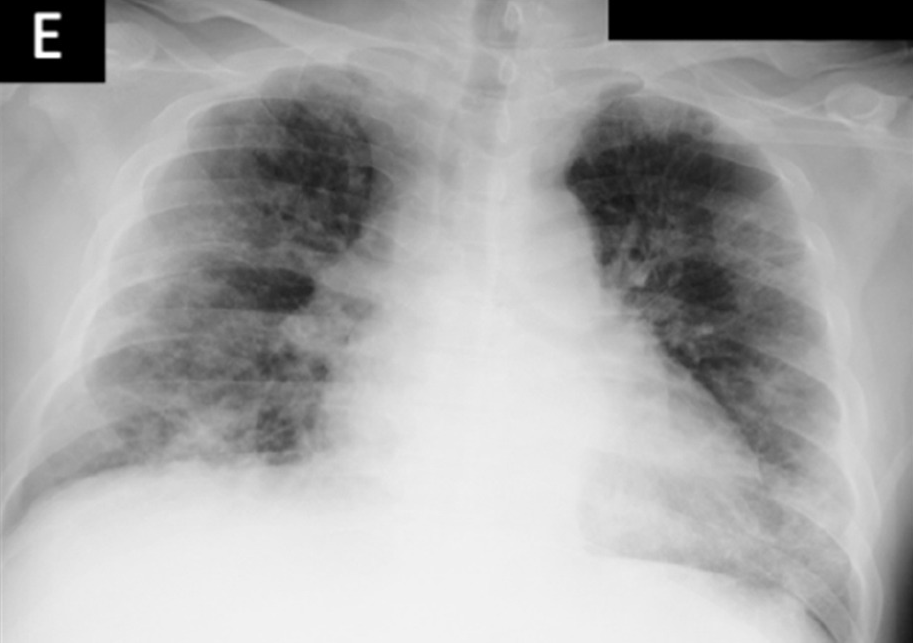}}%
    \caption{Example images from the same patient (\#318) extracted from \citet{10.1016/j.jiac.2020.03.018}. This patient was intubated in the ICU for days 2, 7 and 12.}
    \label{fig:example2}
\end{figure}

\begin{figure}[h]
\centering
\includegraphics[width=0.99\columnwidth]{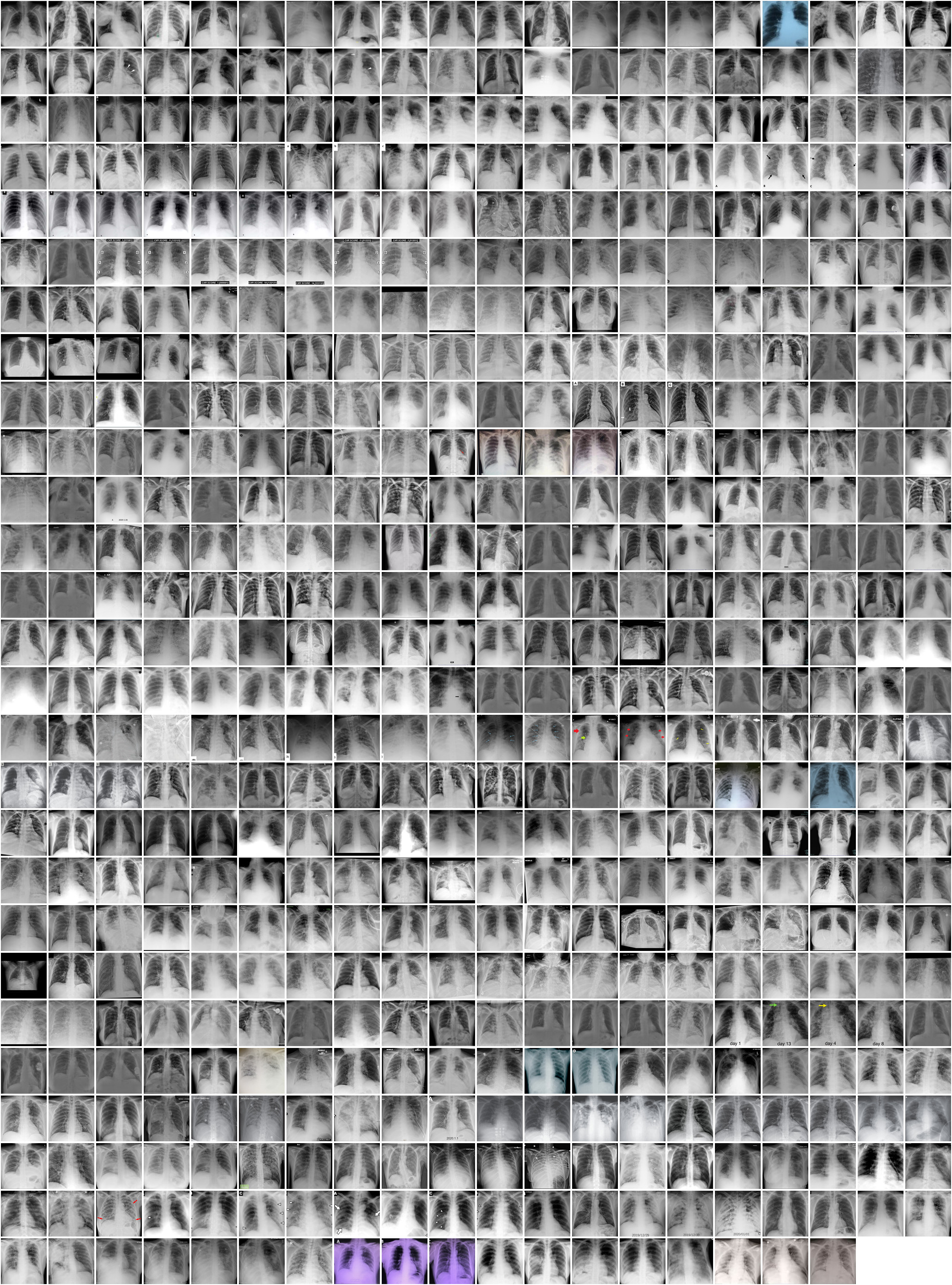}
\caption{A montage of all frontal view (PA, AP, AP Supine) images.}
\label{fig:montage}
\end{figure}

\newpage
\newpage
\section{Scraper details}
\label{apdx:scraper}
Each scraper identifies relevant radiographs using the ``search" feature of the site being scraped, and saves the images together with a csv file of the corresponding metadata. Internally, the scraper uses Selenium to visit each page of the search results, saving metadata and images from any new cases. Images are downloaded at the highest resolution available. Data from each case is converted to a single, interoperable JSON format. Finally, the needed metadata is exported as a csv file.

Currently, the scrapers only retrieve metadata from structured fields that can be easily detected on webpages, such as ``Age", ``Sex", ``Imaging Notes", and ``Clinical Notes." Other clinically important information (such as whether the patient went to the ICU or was intubated) is extracted by human annotators.

Care has been taken to use these websites' resources conservatively and avoid creating a burden on their servers. Crawling is done in a lazy manner, so that no pages are requested until they can be used. Also, within each scraping session, each web page is kept open in an individual browser instance as long as it is needed to reduce re-requested resources .




\newpage

\subsection{Papers where images and clinical data are sourced}
\label{sec:sourcepapers}


\begin{table}[H]
    \centering
    \caption{Papers and counts on specific views}
    \label{tab:my_label}
\resizebox{0.5\textwidth}{!}{%
\begin{tabular}{lrrr}
\toprule
Citation &  PA/AP &  AP Supine &  L \\
\midrule
\cite{10.6084/m9.figshare.12275009}              &     48 &         44 &  0 \\
\cite{10.1148/rg.242035193}                      &     11 &          0 &  0 \\
\cite{10.1148/radiol.2020201160}                 &      9 &          0 &  0 \\
\cite{10.1148/radiol.2282030593}                 &      5 &          0 &  0 \\
\cite{10.1016/j.jfma.2020.02.007}                &      4 &          0 &  0 \\
\cite{10.1016/j.jmii.2020.03.008}                &      4 &          0 &  0 \\
\cite{10.1016/j.radi.2020.04.002}                &      4 &          0 &  0 \\
\cite{10.1056/NEJMoa2001191}                     &      4 &          0 &  3 \\
\cite{10.1056/nejmc2001272}                      &      4 &          0 &  0 \\
\cite{10.1148/ryct.2020200034}                   &      4 &          0 &  0 \\
\cite{10.1007/s00068-020-01364-7}                &      3 &          0 &  0 \\
\cite{10.1007/s00296-020-04584-7}                &      3 &          0 &  0 \\
\cite{10.1016/j.anl.2020.04.002}                 &      3 &          0 &  0 \\
\cite{10.1016/j.chom.2020.03.021}                &      3 &          0 &  0 \\
\cite{10.1016/j.idcr.2020.e00775}                &      3 &          0 &  0 \\
\cite{10.1016/j.medcle.2020.03.004}              &      3 &          0 &  0 \\
\cite{10.1093/cid/ciaa199}                       &      3 &          0 &  0 \\
\cite{10.1148/ryct.2020200033}                   &      3 &          0 &  0 \\
\cite{10.3346/jkms.2020.35.e79}                  &      3 &          0 &  0 \\
\cite{10.3348/kjr.2020.0132}                     &      3 &          0 &  0 \\
\cite{10.4269/ajtmh.20-0203}                     &      3 &          0 &  0 \\
\cite{10.1007/s11547-020-01200-3}                &      2 &          3 &  0 \\
\cite{10.1111/all.14238}                         &      2 &          3 &  0 \\
\cite{10.1016/j.idcr.2020.e00762}                &      2 &          1 &  0 \\
\cite{10.1016/S0140-67362030211-7}             &      2 &          0 &  0 \\
\cite{10.1016/S1473-30992030111-0}             &      2 &          0 &  0 \\
\cite{10.1016/j.cardfail.2020.04.007}            &      2 &          0 &  0 \\
\cite{10.1016/j.crad.2020.04.002}                &      2 &          0 &  0 \\
\cite{10.1016/j.jmii.2020.02.009}                &      2 &          0 &  0 \\
\cite{10.1016/j.jmii.2020.03.003}                &      2 &          0 &  0 \\
\cite{10.1016/j.kint.2020.03.018}                &      2 &          0 &  0 \\
\cite{10.1016/j.seizure.2020.04.009}             &      2 &          0 &  0 \\
\cite{10.1038/s41379-020-0536-x}                 &      2 &          0 &  0 \\
\cite{10.1038/s41591-020-0819-2}                 &      2 &          0 &  0 \\
\cite{10.1056/NEJMc2001573}                      &      2 &          0 &  0 \\
\cite{10.11909/j.issn.1671-5411.2020.04.005}     &      2 &          0 &  0 \\
\cite{10.7759/cureus.7482}                       &      2 &          0 &  0 \\
\cite{10.1007/s11547-020-01203-0}                &      1 &          7 &  0 \\
\cite{10.1016/j.hrcr.2020.04.015}                &      1 &          3 &  0 \\
\cite{10.1016/S2213-26002030076-X}             &      1 &          2 &  0 \\
\cite{10.1016/j.jgar.2020.04.024}                &      1 &          2 &  0 \\
\cite{10.1007/s00270-020-02470-0}                &      1 &          1 &  0 \\
\cite{10.1007/s11547-020-01202-1}                &      1 &          1 &  0 \\
\cite{10.1016/j.clim.2020.108450}                &      1 &          1 &  0 \\
\cite{10.1016/S0140-67362030370-6}             &      1 &          0 &  0 \\
\cite{10.1016/j.healun.2020.04.004}              &      1 &          0 &  0 \\
\cite{10.1016/j.jhin.2020.03.036}                &      1 &          0 &  0 \\
\cite{10.1093/ageing/afaa068}                    &      1 &          0 &  0 \\
\cite{10.1148/cases.20201394}                    &      1 &          0 &  0 \\
\cite{10.1148/cases.20201558}                    &      1 &          0 &  0 \\
\cite{10.1148/cases.20201559}                    &      1 &          0 &  0 \\
\cite{10.1148/cases.20201815}                    &      1 &          0 &  0 \\
\cite{10.1148/radiol.2020200269}                 &      1 &          0 &  0 \\
\cite{10.1148/radiol.2020200274}                 &      1 &          0 &  0 \\
\cite{10.1148/radiol.2020200490}                 &      1 &          0 &  0 \\
\cite{10.1148/ryct.2020200028}                   &      1 &          0 &  0 \\
\cite{10.1186/s12941-020-00358-y}                &      1 &          0 &  0 \\
\cite{10.1186/s40779-020-0233-6}                 &      1 &          0 &  0 \\
\bottomrule
\end{tabular}
}%
\resizebox{0.5\textwidth}{!}{%
\begin{tabular}{lrrr}
\toprule
Citation &  PA/AP &  AP Supine &  L \\
\midrule
\cite{10.3348/kjr.2020.0112}                     &      1 &          0 &  0 \\
\cite{10.4103/1817-1737.69106}                   &      1 &          0 &  0 \\
\cite{10.4274/balkanmedj.galenos.2020.2020.2.15} &      1 &          0 &  0 \\
\cite{10.7150/thno.46465}                        &      1 &          0 &  0 \\
\cite{10.7759/cureus.7473}                       &      1 &          0 &  0 \\
\cite{10.1016/j.jiac.2020.03.018}                &      0 &          5 &  0 \\
\cite{10.1056/NEJMoa2001017}                     &      0 &          2 &  0 \\
\cite{10.1056/NEJMoa2004500}                     &      0 &          2 &  0 \\
\cite{10.7759/cureus.7782}                       &      0 &          2 &  0 \\
\cite{10.1007/s00701-020-04374-x}                &      0 &          1 &  0 \\
\cite{10.1016/j.ajem.2020.04.045}                &      0 &          1 &  0 \\
\cite{10.1016/j.amjoto.2020.102487}              &      0 &          1 &  0 \\
\cite{10.1016/j.bbi.2020.04.077}                 &      0 &          1 &  0 \\
\cite{10.1016/j.chest.2020.04.024}               &      0 &          1 &  0 \\
\cite{10.1016/j.pulmoe.2020.04.012}              &      0 &          1 &  0 \\
\cite{10.2214/AJR.20.23034}                     &      0 &          1 &  0 \\
\cite{10.2214/AJR.20.23034}                      &      0 &          1 &  0 \\
\cite{10.7759/cureus.7352}                       &      0 &          1 &  0 \\
\bottomrule
\end{tabular}
}

\end{table}
\subsection{Output of models on each split}
\label{sec:fullresults}

\begin{table}[H]
\caption{Geographic Extent}
\label{tab:Geographic Extent}
\centering
\resizebox{1\textwidth}{!}{%
\begin{tabular}{lrrrrlll}
\toprule
\# params &  \# test samples &  Correlation &   MAE &   R\textasciicircum 2 &  method &                   name & test\_region \\
\midrule
  1024+1 &            36.0 &         0.73 &  1.46 &  0.30 &  linear &  Intermediate features &        Asia \\
  1024+1 &            41.0 &         0.82 &  1.35 &  0.53 &  linear &  Intermediate features &      Europe \\
     1+1 &            36.0 &         0.00 &  1.92 & -0.15 &  linear &                Baseline prevalence &        Asia \\
     1+1 &            41.0 &         0.00 &  2.36 & -0.51 &  linear &                Baseline prevalence &      Europe \\
    18+1 &            36.0 &         0.73 &  1.25 &  0.47 &  linear &             18 outputs &        Asia \\
    18+1 &            41.0 &         0.86 &  1.04 &  0.71 &  linear &             18 outputs &      Europe \\
     4+1 &            36.0 &         0.78 &  1.17 &  0.58 &  linear &              4 outputs &        Asia \\
     4+1 &            41.0 &         0.86 &  1.06 &  0.66 &  linear &              4 outputs &      Europe \\
     1+1 &            36.0 &         0.77 &  1.15 &  0.57 &  linear &    lung opacity output &        Asia \\
     1+1 &            41.0 &         0.83 &  1.16 &  0.60 &  linear &    lung opacity output &      Europe \\
\bottomrule
\end{tabular}
}
\end{table}

\begin{table}[H]
\caption{Opacity}
\label{tab:opacity}
\centering
\resizebox{1\textwidth}{!}{%
\begin{tabular}{lrrrrlll}
\toprule
\# params &  \# test samples &  Correlation &   MAE &   R\textasciicircum 2 &  method &                   name & test\_region \\
\midrule
  1024+1 &            36.0 &         0.60 &  1.39 & -0.41 &  linear &  Intermediate features &        Asia \\
  1024+1 &            41.0 &         0.76 &  1.02 &  0.22 &  linear &  Intermediate features &      Europe \\
     1+1 &            36.0 &         0.00 &  1.30 & -0.12 &  linear &                Baseline prevalence &        Asia \\
     1+1 &            41.0 &         0.00 &  1.30 & -0.40 &  linear &                Baseline prevalence &      Europe \\
    18+1 &            36.0 &         0.57 &  1.02 &  0.11 &  linear &             18 outputs &        Asia \\
    18+1 &            41.0 &         0.76 &  0.79 &  0.48 &  linear &             18 outputs &      Europe \\
     4+1 &            36.0 &         0.75 &  0.80 &  0.54 &  linear &              4 outputs &        Asia \\
     4+1 &            41.0 &         0.84 &  0.66 &  0.69 &  linear &              4 outputs &      Europe \\
     1+1 &            36.0 &         0.74 &  0.83 &  0.53 &  linear &    lung opacity output &        Asia \\
     1+1 &            41.0 &         0.84 &  0.69 &  0.67 &  linear &    lung opacity output &      Europe \\
\bottomrule
\end{tabular}
}
\end{table}

\begin{table}[H]
\caption{COVID-19}
\label{tab:COVID19}
\centering
\resizebox{1\textwidth}{!}{%
\begin{tabular}{llrrlll}
\toprule
\# params &         \# test samples &  AUPRC &  AUROC &    method &                   name & test\_region \\
\midrule
  1024+1 &   \{True: 39, False: 2\} &   0.95 &   0.50 &  logistic &  Intermediate features &        Asia \\
  1024+1 &    \{False: 3, True: 7\} &   0.70 &   0.50 &  logistic &  Intermediate features &    Americas \\
  1024+1 &    \{False: 4, True: 3\} &   0.60 &   0.75 &  logistic &  Intermediate features &     Oceania \\
  1024+1 &  \{True: 68, False: 14\} &   0.84 &   0.55 &  logistic &  Intermediate features &      Europe \\
     1+1 &   \{True: 39, False: 2\} &   0.95 &   0.50 &  logistic &    Baseline prevalence &        Asia \\
     1+1 &    \{False: 3, True: 7\} &   0.70 &   0.50 &  logistic &    Baseline prevalence &    Americas \\
     1+1 &    \{False: 4, True: 3\} &   0.43 &   0.50 &  logistic &    Baseline prevalence &     Oceania \\
     1+1 &  \{True: 68, False: 14\} &   0.83 &   0.50 &  logistic &    Baseline prevalence &      Europe \\
    18+1 &   \{True: 39, False: 2\} &   0.95 &   0.49 &  logistic &             18 outputs &        Asia \\
    18+1 &    \{False: 3, True: 7\} &   0.70 &   0.50 &  logistic &             18 outputs &    Americas \\
    18+1 &    \{False: 4, True: 3\} &   0.60 &   0.75 &  logistic &             18 outputs &     Oceania \\
    18+1 &  \{True: 68, False: 14\} &   0.86 &   0.59 &  logistic &             18 outputs &      Europe \\
     4+1 &   \{True: 39, False: 2\} &   0.95 &   0.50 &  logistic &              4 outputs &        Asia \\
     4+1 &    \{False: 3, True: 7\} &   0.70 &   0.50 &  logistic &              4 outputs &    Americas \\
     4+1 &    \{False: 4, True: 3\} &   0.43 &   0.50 &  logistic &              4 outputs &     Oceania \\
     4+1 &  \{True: 68, False: 14\} &   0.83 &   0.52 &  logistic &              4 outputs &      Europe \\
     1+1 &   \{True: 39, False: 2\} &   0.95 &   0.50 &  logistic &    lung opacity output &        Asia \\
     1+1 &    \{False: 3, True: 7\} &   0.70 &   0.50 &  logistic &    lung opacity output &    Americas \\
     1+1 &    \{False: 4, True: 3\} &   0.43 &   0.50 &  logistic &    lung opacity output &     Oceania \\
     1+1 &  \{True: 68, False: 14\} &   0.83 &   0.51 &  logistic &    lung opacity output &      Europe \\
 5017801 &   \{True: 39, False: 2\} &   0.94 &   0.42 &       MLP &     Image pixels (MLP) &        Asia \\
 5017801 &   \{True: 39, False: 2\} &   0.95 &   0.50 &       MLP &     Image pixels (MLP) &        Asia \\
 5017801 &   \{True: 39, False: 2\} &   0.95 &   0.50 &       MLP &     Image pixels (MLP) &        Asia \\
 5017801 &    \{False: 3, True: 7\} &   0.67 &   0.43 &       MLP &     Image pixels (MLP) &    Americas \\
 5017801 &    \{False: 3, True: 7\} &   0.65 &   0.36 &       MLP &     Image pixels (MLP) &    Americas \\
 5017801 &    \{False: 3, True: 7\} &   0.70 &   0.50 &       MLP &     Image pixels (MLP) &    Americas \\
 5017801 &    \{False: 4, True: 3\} &   0.43 &   0.50 &       MLP &     Image pixels (MLP) &     Oceania \\
 5017801 &    \{False: 4, True: 3\} &   0.43 &   0.50 &       MLP &     Image pixels (MLP) &     Oceania \\
 5017801 &    \{False: 4, True: 3\} &   0.43 &   0.50 &       MLP &     Image pixels (MLP) &     Oceania \\
 5017801 &  \{True: 68, False: 14\} &   0.83 &   0.49 &       MLP &     Image pixels (MLP) &      Europe \\
 5017801 &  \{True: 68, False: 14\} &   0.86 &   0.60 &       MLP &     Image pixels (MLP) &      Europe \\
 5017801 &  \{True: 68, False: 14\} &   0.83 &   0.49 &       MLP &     Image pixels (MLP) &      Europe \\
\bottomrule
\end{tabular}
}
\end{table}

\begin{table}[H]
\caption{Viral or Bacterial}
\label{tab:ViralorBacterial}
\centering
\resizebox{1\textwidth}{!}{%
\begin{tabular}{llrrllll}
\toprule
\# params &        \# test samples &  AUPRC &  AUROC &    method &                   Features & \multicolumn{2}{l}{test\_region} \\
\midrule
 1024+1 &   \{False: 3, True: 3\} &   0.75 &   0.83 &  logistic &  Intermediate features &     Oceania \\
  1024+1 &  \{True: 72, False: 6\} &   0.93 &   0.57 &  logistic &  Intermediate features &      Europe \\
     1+1 &   \{False: 3, True: 3\} &   0.50 &   0.50 &  logistic &    Baseline prevalence &     Oceania \\
     1+1 &  \{True: 72, False: 6\} &   0.92 &   0.50 &  logistic &    Baseline prevalence &      Europe \\
    18+1 &   \{False: 3, True: 3\} &   1.00 &   1.00 &  logistic &             18 outputs &     Oceania \\
    18+1 &  \{True: 72, False: 6\} &   0.94 &   0.64 &  logistic &             18 outputs &      Europe \\
     4+1 &   \{False: 3, True: 3\} &   0.50 &   0.50 &  logistic &              4 outputs &     Oceania \\
     4+1 &  \{True: 72, False: 6\} &   0.93 &   0.54 &  logistic &              4 outputs &      Europe \\
     1+1 &   \{False: 3, True: 3\} &   0.50 &   0.50 &  logistic &    lung opacity output &     Oceania \\
     1+1 &  \{True: 72, False: 6\} &   0.94 &   0.61 &  logistic &    lung opacity output &      Europe \\
 5017801 &   \{False: 3, True: 3\} &   0.50 &   0.50 &       MLP &     Image pixels (MLP) &     Oceania \\
 5017801 &   \{False: 3, True: 3\} &   0.50 &   0.50 &       MLP &     Image pixels (MLP) &     Oceania \\
 5017801 &   \{False: 3, True: 3\} &   0.50 &   0.50 &       MLP &     Image pixels (MLP) &     Oceania \\
 5017801 &  \{True: 72, False: 6\} &   0.92 &   0.50 &       MLP &     Image pixels (MLP) &      Europe \\
 5017801 &  \{True: 72, False: 6\} &   0.92 &   0.50 &       MLP &     Image pixels (MLP) &      Europe \\
 5017801 &  \{True: 72, False: 6\} &   0.92 &   0.48 &       MLP &     Image pixels (MLP) &      Europe \\
\bottomrule
\end{tabular}
}
\end{table}

\begin{table}[H]
\caption{Survival prediction}
\label{tab:Survivalprediction}
\centering
\resizebox{1\textwidth}{!}{%
\begin{tabular}{llrrlll}
\toprule
\# params &        \# test samples &  AUPRC &  AUROC &    method &                   Features & test\_region \\
\midrule
  1024+1 &  \{True: 20, False: 3\} &   0.87 &   0.50 &  logistic &  Intermediate features &        Asia \\
  1024+1 &   \{False: 1, True: 4\} &   0.80 &   0.50 &  logistic &  Intermediate features &    Americas \\
  1024+1 &  \{True: 19, False: 3\} &   0.86 &   0.50 &  logistic &  Intermediate features &      Europe \\
     1+1 &  \{True: 20, False: 3\} &   0.87 &   0.50 &  logistic &    Baseline prevalence &        Asia \\
     1+1 &   \{False: 1, True: 4\} &   0.80 &   0.50 &  logistic &    Baseline prevalence &    Americas \\
     1+1 &  \{True: 19, False: 3\} &   0.86 &   0.50 &  logistic &    Baseline prevalence &      Europe \\
    18+1 &  \{True: 20, False: 3\} &   0.86 &   0.45 &  logistic &             18 outputs &        Asia \\
    18+1 &   \{False: 1, True: 4\} &   0.80 &   0.50 &  logistic &             18 outputs &    Americas \\
    18+1 &  \{True: 19, False: 3\} &   0.85 &   0.45 &  logistic &             18 outputs &      Europe \\
     4+1 &  \{True: 20, False: 3\} &   0.87 &   0.50 &  logistic &              4 outputs &        Asia \\
     4+1 &   \{False: 1, True: 4\} &   0.80 &   0.50 &  logistic &              4 outputs &    Americas \\
     4+1 &  \{True: 19, False: 3\} &   0.89 &   0.61 &  logistic &              4 outputs &      Europe \\
     1+1 &  \{True: 20, False: 3\} &   0.87 &   0.50 &  logistic &    lung opacity output &        Asia \\
     1+1 &   \{False: 1, True: 4\} &   0.80 &   0.50 &  logistic &    lung opacity output &    Americas \\
     1+1 &  \{True: 19, False: 3\} &   0.90 &   0.64 &  logistic &    lung opacity output &      Europe \\
 5017801 &  \{True: 20, False: 3\} &   0.87 &   0.50 &       MLP &     Image pixels (MLP) &        Asia \\
 5017801 &  \{True: 20, False: 3\} &   0.87 &   0.50 &       MLP &     Image pixels (MLP) &        Asia \\
 5017801 &  \{True: 20, False: 3\} &   0.87 &   0.50 &       MLP &     Image pixels (MLP) &        Asia \\
 5017801 &   \{False: 1, True: 4\} &   0.80 &   0.50 &       MLP &     Image pixels (MLP) &    Americas \\
 5017801 &   \{False: 1, True: 4\} &   0.80 &   0.50 &       MLP &     Image pixels (MLP) &    Americas \\
 5017801 &   \{False: 1, True: 4\} &   0.80 &   0.50 &       MLP &     Image pixels (MLP) &    Americas \\
 5017801 &  \{True: 19, False: 3\} &   0.86 &   0.50 &       MLP &     Image pixels (MLP) &      Europe \\
 5017801 &  \{True: 19, False: 3\} &   0.86 &   0.50 &       MLP &     Image pixels (MLP) &      Europe \\
 5017801 &  \{True: 19, False: 3\} &   0.86 &   0.50 &       MLP &     Image pixels (MLP) &      Europe \\
\bottomrule
\end{tabular}
}
\end{table}

\begin{table}[H]
\caption{ICU Stay}
\label{tab:ICUStay}
\centering
\resizebox{1\textwidth}{!}{%
\begin{tabular}{llrrlll}
\toprule
\# params &         \# test samples &  AUPRC &  AUROC &    method &                   Features & test\_region \\
\midrule
  1024+1 &    \{False: 9, True: 1\} &   0.25 &   0.83 &  logistic &  Intermediate features &        Asia \\
  1024+1 &  \{False: 15, True: 13\} &   0.55 &   0.58 &  logistic &  Intermediate features &      Europe \\
     1+1 &    \{False: 9, True: 1\} &   0.10 &   0.50 &  logistic &    Baseline prevalence &        Asia \\
     1+1 &  \{False: 15, True: 13\} &   0.46 &   0.50 &  logistic &    Baseline prevalence &      Europe \\
    18+1 &    \{False: 9, True: 1\} &   0.33 &   0.89 &  logistic &             18 outputs &        Asia \\
    18+1 &  \{False: 15, True: 13\} &   0.67 &   0.74 &  logistic &             18 outputs &      Europe \\
     4+1 &    \{False: 9, True: 1\} &   0.10 &   0.44 &  logistic &              4 outputs &        Asia \\
     4+1 &  \{False: 15, True: 13\} &   0.61 &   0.66 &  logistic &              4 outputs &      Europe \\
     1+1 &    \{False: 9, True: 1\} &   0.10 &   0.33 &  logistic &    lung opacity output &        Asia \\
     1+1 &  \{False: 15, True: 13\} &   0.46 &   0.50 &  logistic &    lung opacity output &      Europe \\
 5017801 &    \{False: 9, True: 1\} &   0.11 &   0.56 &       MLP &     Image pixels (MLP) &        Asia \\
 5017801 &    \{False: 9, True: 1\} &   0.10 &   0.50 &       MLP &     Image pixels (MLP) &        Asia \\
 5017801 &    \{False: 9, True: 1\} &   0.10 &   0.50 &       MLP &     Image pixels (MLP) &        Asia \\
 5017801 &  \{False: 15, True: 13\} &   0.46 &   0.30 &       MLP &     Image pixels (MLP) &      Europe \\
 5017801 &  \{False: 15, True: 13\} &   0.46 &   0.27 &       MLP &     Image pixels (MLP) &      Europe \\
 5017801 &  \{False: 15, True: 13\} &   0.46 &   0.27 &       MLP &     Image pixels (MLP) &      Europe \\
\bottomrule
\end{tabular}
}
\end{table}

\begin{table}[H]
\caption{Intubated}
\label{tab:Intubated}
\centering
\resizebox{1\textwidth}{!}{%
\begin{tabular}{llrrlll}
\toprule
\# params &        \# test samples &  AUPRC &  AUROC &    method &                   Features & test\_region \\
\midrule
  1024+1 &  \{False: 10, True: 2\} &   0.25 &   0.65 &  logistic &  Intermediate features &        Asia \\
  1024+1 &   \{False: 8, True: 7\} &   0.54 &   0.57 &  logistic &  Intermediate features &      Europe \\
     1+1 &  \{False: 10, True: 2\} &   0.17 &   0.50 &  logistic &    Baseline prevalence &        Asia \\
     1+1 &   \{False: 8, True: 7\} &   0.47 &   0.50 &  logistic &    Baseline prevalence &      Europe \\
    18+1 &  \{False: 10, True: 2\} &   0.17 &   0.30 &  logistic &             18 outputs &        Asia \\
    18+1 &   \{False: 8, True: 7\} &   0.53 &   0.61 &  logistic &             18 outputs &      Europe \\
     4+1 &  \{False: 10, True: 2\} &   0.17 &   0.45 &  logistic &              4 outputs &        Asia \\
     4+1 &   \{False: 8, True: 7\} &   0.45 &   0.45 &  logistic &              4 outputs &      Europe \\
     1+1 &  \{False: 10, True: 2\} &   0.17 &   0.50 &  logistic &    lung opacity output &        Asia \\
     1+1 &   \{False: 8, True: 7\} &   0.47 &   0.50 &  logistic &    lung opacity output &      Europe \\
 5017801 &  \{False: 10, True: 2\} &   0.17 &   0.50 &       MLP &     Image pixels (MLP) &        Asia \\
 5017801 &  \{False: 10, True: 2\} &   0.17 &   0.50 &       MLP &     Image pixels (MLP) &        Asia \\
 5017801 &  \{False: 10, True: 2\} &   0.18 &   0.55 &       MLP &     Image pixels (MLP) &        Asia \\
 5017801 &   \{False: 8, True: 7\} &   0.50 &   0.56 &       MLP &     Image pixels (MLP) &      Europe \\
 5017801 &   \{False: 8, True: 7\} &   0.47 &   0.38 &       MLP &     Image pixels (MLP) &      Europe \\
 5017801 &   \{False: 8, True: 7\} &   0.47 &   0.50 &       MLP &     Image pixels (MLP) &      Europe \\
\bottomrule
\end{tabular}
}
\end{table}

\end{document}